\documentclass[aps,preprint,showpacs,groupedaddress,floatfix]{revtex4}

\usepackage{graphicx}
\usepackage{dcolumn}
\usepackage{bm}
\usepackage{rotating}
\usepackage{xcolor}

\begin{document}

\bibliographystyle{prsty}
\title{Low-energy isovector and isoscalar dipole response in neutron-rich nuclei}
\author{D. Vretenar$^1$}
\author{Y. F. Niu$^{1,2}$}
\author{N. Paar$^1$}
\author{J. Meng$^{2,3,4}$}
 \affiliation{$^1$Physics Department, Faculty of Science, University of Zagreb, Croatia}
 \affiliation{$^2$State Key Laboratory for Nuclear Physics and Technology, School of Physics, Peking University, Beijing 100871, China}
 \affiliation{$^{3}$School of Physics and Nuclear Energy Engineering, Beihang University, Beijing 100191, China}
 \affiliation{$^4$ Department of Physics, University of Stellenbosch, Stellenbosch 7602, South Africa}

\date{\today}
\begin{abstract}
The self-consistent random phase approximation (RPA), based on the
framework of relativistic energy density functionals, is employed in
the study of isovector and isoscalar dipole response in $^{68}$Ni,
$^{132}$Sn, and $^{208}$Pb. The evolution of pygmy dipole states
(PDS) in the region of low excitation energies is analyzed as a
function of the density-dependence of the symmetry energy for a set
of relativistic effective interactions. The occurrence of PDS is
predicted in the response to both the isovector and isoscalar dipole
operators, and its strength is enhanced with the increase of the
symmetry energy at saturation and the slope of the symmetry energy.
In both channels the PDS exhausts a relatively small fraction of the
energy-weighted sum rule but a much larger percentage of the inverse
energy-weighted sum rule. For the isovector dipole operator the
reduced transition probability $B(E1)$ of the PDS  is generally
small because of pronounced cancellation of neutron and proton partial
contributions. The isoscalar reduced transition amplitude is
predominantly determined by neutron particle-hole configurations,
most of which add coherently, and
this results in a collective response of the PDS to the isoscalar
dipole operator.
\end{abstract}
\pacs{ 21.60.Jz, 24.30.Cz, 24.30.Gd, 21.65.Ef, 21.30.Fe}
\maketitle
\date{today}
\section{Introduction}
Studies of collective excitations play a crucial role in our understanding of
complex properties of nuclei. The multipole response of unstable nuclei and, in particular,
a possible occurrence of new exotic modes of excitation in weakly bound nuclear systems
presents a rapidly growing field of research~\cite{PVKC.07,Paa.10}. During the last decade a number of experimental studies have been performed on the low-energy electric dipole response in neutron-rich
medium-heavy and heavy nuclei \cite{Rye.02,Zil.02,Hart.04,Adr.05,Sav.06,Vol.06,Kli.07,Kli.07a,
Oze.07,Sav.08,Sch.08,End.09,Wie.09,Ton.10,End.10,Sav.11,Isa.11,Tam.11}. The currently available
data, however, are not sufficient to completely determine the nature of observed excitations,
and the interpretation of the dynamics of low-energy E1 strength is under discussion.

The structure and dynamics of low-energy dipole strength, also referred to as pygmy dipole
state (PDS) or resonance, has extensively been investigated
using a variety of theoretical approaches and models \cite{PVKC.07}.
Recent studies have made use of the Hartree-Fock (HF) plus
random phase approximation (RPA) \cite{Peru.05,Lan.09,Car.10,RN.10,Lan.11,Ina.11,Yuk.12,RM.11},
the self-consistent RPA based on the AV18 nucleon-nucleon interaction and
phenomenological three-body contact terms~\cite{Pan.11},
the Hartree-Fock-Bogoliubov (HFB) model plus quasiparticle
(Q)RPA \cite{Mat.05,Ter.06,Yosh.08,Li.08,Yosh.09,Eba.10,Mart.11},
quasiparticle RPA plus phonon coupling~\cite{Sar.04},
the quasiparticle-phonon model~\cite{Tso.08}, the
quasiparticle phonon model including complex configurations of up to three
phonons \cite{Sav.08,End.10,Sav.11}, the second RPA (SRPA) \cite{GGC.11},
the relativistic RPA~\cite{Pie.06,Lia.07,Liang.08,Pie.11} and
QRPA~\cite{PNVR.05,Paa.09,PKR.09,Dao.11}, the relativistic quasiparticle
time blocking approximation~\cite{Lit.07,Lit.08,Lit.09,Lit.10}, and the semiclassical
Landau-Vlasov approach \cite{Ur.11,BFCT.11}.

In addition to the fact that pygmy states present an intrinsically interesting structure phenomenon
in neutron-rich nuclei, it has been suggested that they might constrain the radius of
the neutron distribution in medium-heavy and heavy nuclei \cite{Kli.07}, and provide
information about the density dependence of the symmetry energy \cite{Pie.06}.
Prompted by recent experiments that have reported accurate data on
the distribution of PDS in medium heavy-nuclei \cite{Adr.05,Sav.06,Vol.06,Kli.07,Kli.07a,
Oze.07,Sav.08,Sch.08,End.09,Wie.09,Ton.10,End.10,Sav.11}, and by the first
model-independent determination of the neutron-skin thickness in $^{208}$Pb
with parity-violating elastic electron scattering \cite{PREX}, in the last two years
several theoretical studies have explored the relation between the PDS, the
density dependence of the symmetry energy, and the neutron skin \cite{RN.10,Car.10,Pie.11,RM.11}.
These studies are based on the microscopic framework of nuclear energy density functionals (EDF)
plus the (Q)RPA.

Starting from a representative set of Skyrme effective forces and meson-exchange effective
Lagrangians, Carbone et al. \cite{Car.10} performed an RPA analysis of the correlation between
the density dependence of the nuclear symmetry energy, the neutron skin, and the percentage of
energy-weighted sum rule (EWSR) exhausted by the PDS in $^{68}$Ni and $^{132}$Sn. In
comparison with available data it was possible to constrain the value of the derivative of the
symmetry energy at saturation, and use this constraint to determine the neutron-skin radii of
$^{68}$Ni, $^{132}$Sn, and $^{208}$Pb. In contrast to this result, Reinhard and
Nazarewicz  \cite{RN.10}, using the covariance analysis to identify observables and
pseudo-observables that correlate with the neutron skin,
suggested that the neutron skin of $^{208}$Pb is strongly correlated with the dipole polarizability
but very weakly correlated with the low-energy electric dipole strength. This finding has
recently been challenged by Piekarewicz \cite{Pie.11} in an analysis of the distribution of
electric dipole strength in $^{68}$Ni using a relativistic RPA with a set of effective interactions
that predict significantly different values for the neutron-skin thickness of $^{208}$Pb.
The results suggest a strong correlation between the dipole polarizability of $^{68}$Ni and the
neutron-skin thickness of $^{208}$Pb, but also a correlation just as strong between the neutron-skin
thickness of $^{208}$Pb and the fraction of the dipole polarizability exhausted by the pygmy
dipole strength. In a very recent study performed using the self-consistent Skyrme HF plus
RPA approach, X. Roca-Maza et al. \cite{RM.11} have analyzed the isospin character,
the degree of collectivity, and the sensitivity to the slope of the nuclear symmetry energy,
of the low-energy dipole response in $^{68}$Ni, $^{132}$Sn, and $^{208}$Pb. It has been
shown that both the isoscalar and isovector strength functions display a low-energy peak
that is enhanced and shifted to higher excitation energies with increasing values of the slope of
the symmetry energy at saturation. The degree of collectivity associated with the RPA state(s)
contributing to this peak differs in the isoscalar and isovector channels. Much more
collectivity in the PDS is predicted in the response to the isoscalar dipole operator.

In this work we perform an analysis similar to that of Ref.~\cite{RM.11}, but using a more systematic
set of effective nuclear interactions. Namely, to analyze the model dependence of the predicted PDS,
X. Roca-Maza et al.~\cite{RM.11} employed three different Skyrme parameter sets: SGII, SLy5 and SkI3.
These interactions span a broad range of values of the slope of the nuclear symmetry energy at
saturation, but they also differ in other characteristics in a nonsystematic way. A consistent set of
effective interactions was used by Piekarewicz in Ref.~\cite{Pie.11} to analyze the distribution of
PDS, but only for $^{68}$Ni and only in the isovector channel.
\section{Theoretical framework}

The present analysis employs the fully self-consistent relativistic
random phase approximation (RRPA) based on the framework of
relativistic energy density functionals \cite{NVR.02}. In the
relativistic mean-field (RMF) +
RPA model effective interactions are implemented in a fully
consistent way: effective Lagrangians with density-dependent
meson-nucleon couplings are employed \cite{NVFR.02,LNVR.05} and the
same interactions are used both in the RMF equations that determine the
ground state, and in the matrix equations of the RRPA. The full set
of RRPA equations is solved by diagonalization. The results are
excitation energies $E_{\lambda}$ and the corresponding forward- and
backward-going amplitudes, $X^{\lambda}$ and $Y^{\lambda}$,
respectively, that are used to evaluate the reduced transition
probability from an excited state $\vert J \lambda \rangle$ to the
ground state:
\begin{eqnarray}
B^T(EJ) & = & \frac{1}{2J_{i}+1}
\bigg\vert \sum_{\mu\mu'} \bigg\{ X^{\lambda, J}_{\mu\mu'} \langle
\mu || \hat{Q}^T_J || \mu' \rangle \nonumber \\
& + &~(-1)^{j_{\mu}-j_{\mu'}+J} \, Y^{\lambda, J}_{\mu\mu'}
\, \langle \mu' || \hat{Q}^T_J || \mu \rangle \,
\bigg\}
\bigg\vert ^2 \quad ,
\label{BE}
\end{eqnarray}
where $\mu$ and $\mu'$ denote single-nucleon states. Discrete
spectra are averaged with a Lorentzian distribution of arbitrary
width (1.5 MeV in the
present calculation). The electric E1 response is calculated for the
isovector dipole operator:
\begin{equation}
\label{iv-dip}
\hat{Q}_{1 \mu}^{T=1} \ = \frac{N}{N+Z}\sum^{Z}_{p=1} r_{p}Y_{1 \mu}( {\hat r}_{p})
- \frac{Z}{N+Z}\sum^{N}_{n=1} r_{n}Y_{1 \mu}( {\hat r}_{n}) \; ,
\end{equation}
and the isoscalar dipole operator:
\begin{equation}
\label{is-dip}
\hat{Q}_{1 \mu}^{T=0} \ = \sum^{A}_{i=1} r^3_{i}Y_{1 \mu}( {\hat r}_{i})
- \eta\sum^{A}_{i=1} r_{i}Y_{1 \mu}( {\hat r}_{i}) \; .
\end{equation}
The inclusion of the second term in the isoscalar operator Eq.~(\ref{is-dip}),
with $\eta=5 \langle r^2 \rangle /3$, ensures that the corresponding strength
distribution does not contain spurious components associated to the center-of-mass
motion \cite{VWR.00}. The strength function reads
\begin{equation}
S(E) = \sum_\nu \vert \langle \nu \vert\vert \hat Q_{J}^{T}
\vert\vert 0 \rangle \vert^2 \delta(E-E_\nu) \;.
\label{strength}
\end{equation}
$E_\nu$ is the energy of the RPA state $\vert \nu \rangle$, and
the moments of the strength distribution are defined as
\begin{equation}
m_k = \int dE\ E^k S(E) = \sum_\nu E_\nu^k ~ \vert \langle \nu \vert\vert
\hat Q_{J}^{T} \vert\vert  0 \rangle \vert^2 \;.
\end{equation}

In the following we analyze the occurrence and structure of the PDS in the isovector
and isoscalar dipole response of $^{68}$Ni, $^{132}$Sn, and $^{208}$Pb, in
relation with the density dependence of the symmetry energy. In linear order with
respect to the nuclear matter density $\rho$, the symmetry energy $S(\rho)$ is
determined by its value at saturation density $S(\rho_0) \equiv a_4$, and by the
derivative at saturation density:
\begin{equation}
S^\prime(\rho)\vert_{\rho=\rho_0} \equiv {L\over 3 \rho_0} \;,
\label{slope}
\end{equation}
and this relation defines the ``slope" parameter $L$. Using data on the percentage of
the EWSR associated with the PDS in $^{68}$Ni \cite{Wie.09} and $^{132}$Sn \cite{Kli.07},
Carbone et al. \cite{Car.10} constrained the value of the slope parameter:
$L = 64.8 \pm 15.7$ MeV, in accordance with values previously determined with
different types of analyses and/or other methods based on nuclear structure and
heavy-ion experiments. Here we employ the framework of relativistic energy
density functionals represented by effective Lagrangians with density-dependent
meson-nucleon vertex functions. The parameters of the very successful effective
interactions DD-ME1 \cite{NVFR.02} and DD-ME2 \cite {LNVR.05}, in particular,
were adjusted simultaneously to empirical properties of symmetric and asymmetric
nuclear matter, and to binding energies and charge radii of twelve spherical nuclei.
Data on excitation energies of isoscalar monopole (ISGMR) and isovector dipole
giant resonances (IVGDR) were also used to determine the compressibility
modulus and asymmetry energy at saturation, as well as available data on differences
between neutron and proton radii in Sn isotopes and $^{208}$Pb. Numerous calculations
have shown that these interactions provide accurate results for ground-state properties
of spherical and deformed nuclei, as well as excitation energies of giant resonances.
Pertinent to the present analysis, the relativistic RPA with the DD-ME2 effective
interaction predicts the dipole polarizability:
\begin{equation}
\alpha_D = {{8 \pi}\over 9} e^2~m_{-1}
\label{dip-pol}
\end{equation}
(directly proportional to the inverse energy-weighted moment $m_{-1}$)
for $^{208}$Pb: 20.8 fm$^3$, in very good
agreement with the recently measured value: $\alpha_D = (20.1\pm 0.6)$
fm$^3$ \cite{Tam.11}.
\section{Results}
\begin{figure}
\centering

\includegraphics[scale=0.30,angle=0]{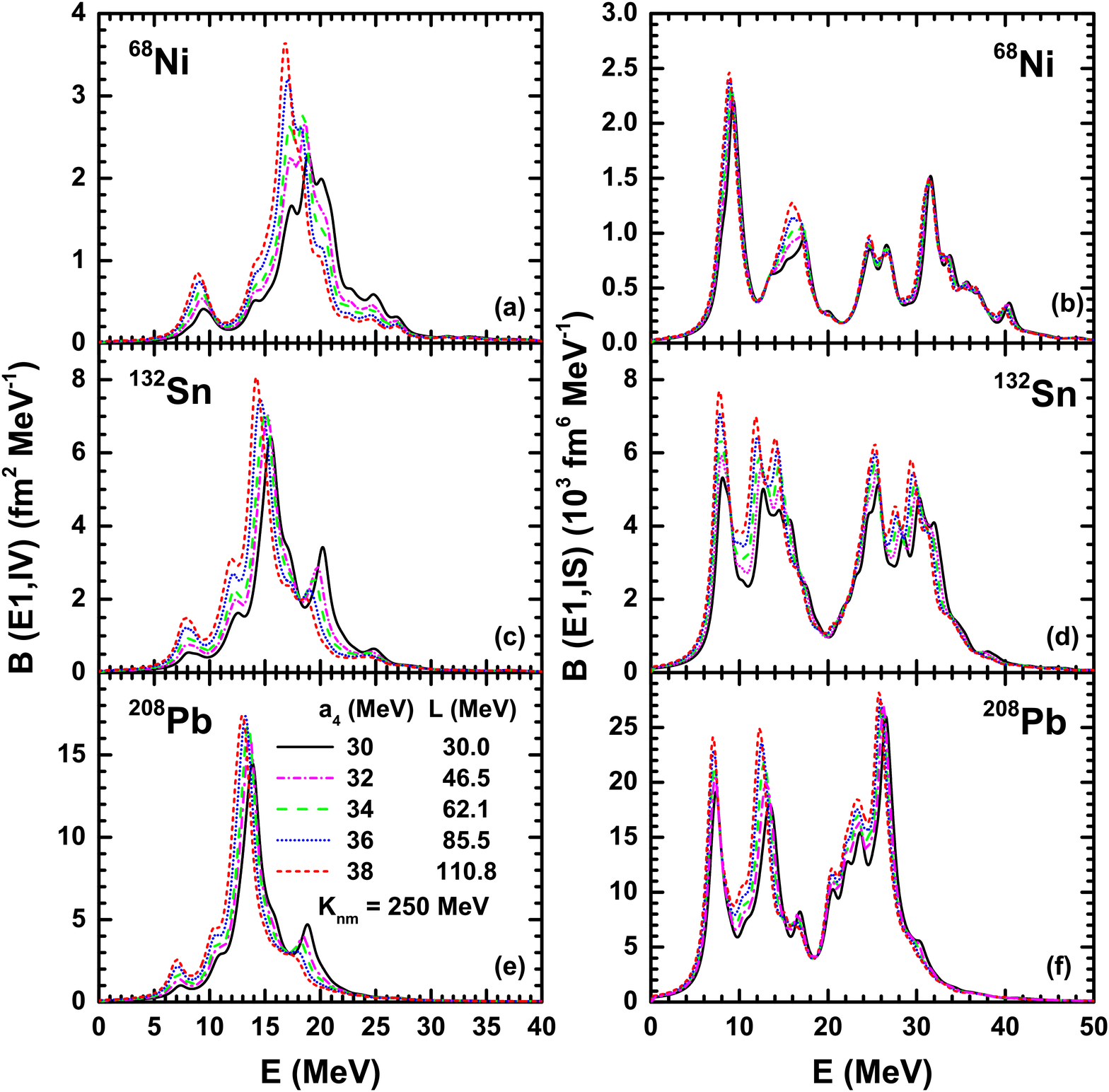}

\caption{(Color online) The relativistic RPA isovector (left column) and
isocalar (right column) dipole strength distributions in $^{68}$Ni,
$^{132}$Sn, and $^{208}$Pb, calculated with a set of five effective
interactions that differ in the value of the symmetry energy at
saturation ($a_4$) and the corresponding slope parameter $L$. }
\label{E1-IV&IS}
\end{figure}

In an earlier study \cite{VNR.03} we used the relativistic RPA with density-dependent
meson-nucleon effective interactions to provide a microscopic estimate of the
nuclear matter compressibility and symmetry energy in relativistic mean-field models.
Starting from the parameterization DD-ME1  \cite{NVFR.02}, three
families of interactions with nuclear matter incompressibility $K_{\rm nm} =$ 230, 250, and 270 MeV
were constructed. For each value of $K_{\rm nm}$ five interactions were
adjusted with $a_4 =$ 30, 32, 34, 36 and 38 MeV, respectively. For particular values of
$K_{\rm nm}$ and $a_4$, the remaining parameters of
these interactions were fine tuned to empirical properties of nuclear matter,
and to the binding energies and charge radii of ten spherical nuclei.
It was noted that, to reproduce the
empirical masses of $N\neq Z$ nuclei, larger values of $a_4$ necessitate
an increase of the slope parameter $L$ of the symmetry energy. These
effective interactions were used in Ref.~\cite{Kli.07} to study a possible
correlation between the observed PDS in $^{130,132}$Sn and the corresponding
values for the neutron skin thickness. In addition to DD-ME2 ($K_{\rm nm} =$ 251 MeV),
the set of effective interactions with $K_{\rm nm} =$ 250 MeV and
$a_4 =$ 30, 32, 34, 36 and 38 MeV will here be used to study PDS in the isovector
and isoscalar dipole response of spherical closed-shell nuclei.

For this set of effective interactions, in
Fig.~\ref{E1-IV&IS} we display the total isovector (IV) and
isoscalar (IS) RRPA dipole strength functions for $^{68}$Ni,
$^{132}$Sn, and $^{208}$Pb. The IV dipole response in all three
nuclei is, of course, dominated by the collective giant dipole
resonance (GDR) peaked in the high-energy region around 15 MeV. As
it was already shown in Refs. \cite{Rei.99,NVR.02}, as a result of
the increase of $L$ with $a_4$ the excitation energy of the IV GDR
decreases with increasing $S(\rho_{0}) \equiv a_4$, because this
increase implies a reduction of the symmetry energy at low densities
characteristic for surface modes. In addition, one notices an
enhancement of E1 strength in the low-energy region below 10 MeV.
This pygmy dipole strength (PDS) is very sensitive to the density
dependence of the symmetry energy, and is strongly enhanced by the
increase of $L$ and $a_4$. As shown in
Ref.~\cite{RM.11}, the same pygmy states are also present in the IS
strength functions, and this is a clear evidence of the mixed
isovector-isoscalar nature of the PDS. The isoscalar E1 strength
distributions display a characteristic bimodal structure with two
broad components: one in the low-energy region close to the IV GDR
($\approx 2 \hbar \omega$), and the other at higher energy close to
the electric octupole resonance ($\approx 3 \hbar \omega$).
Theoretical analyses have shown that the high-energy component
represents compressional vibrations \cite{Colo.00,VWR.00,PVNR.06}.
The high-energy IS GDR is a second order effect, built on $3 \hbar
\omega$ or higher configurations, and corresponds to a compression
wave traveling back and forth through the nucleus along a definite
direction. Some states comprising the broad structure in the
low-energy region correspond to vortical nuclear flow associated
with the toroidal dipole moment \cite{BMS.93,VPRN.02,Mis.06,Ur.11}.
However, as pointed out in a study of the interplay between
compressional and vortical nuclear currents \cite{Mis.06}, a strong
mixing between compressional and vorticity vibrations in the
isoscalar E1 states can be expected up to the highest excitation
energies in the region $\approx 3 \hbar \omega$. Finally, the lowest
peaks in the isoscalar strength functions are associated to the PDS.
A very interesting result, also pointed out in Ref.~\cite{RM.11}, is
that relatively to the corresponding GDR, the PDS is much more
pronounced in the isoscalar channel. In fact, for $^{68}$Ni the
pygmy state has the largest $B(E1)$ value among the isoscalar
states, and the IS strength function, except the structure around 15
MeV, is not sensitive to the variation of the density dependence of
the symmetry energy. One expects more isospin mixing in the two
heavier nuclei (see also the analysis of Ref.~\cite{Paa.09}), and
this is reflected in the enhancement of the $B(E1)$ values with the
increase of $L$ and $a_4$, for all states in the response of
$^{132}$Sn and $^{208}$Pb to the isoscalar dipole operator
Eq.~(\ref{is-dip}). This enhancement is more pronounced below the IS
GDR, that is, in the region around 15 MeV where one expects mixing
with the IV GDR, and for the PDS.

\begin{figure}
\centerline{
\includegraphics[scale=0.37,angle=0]{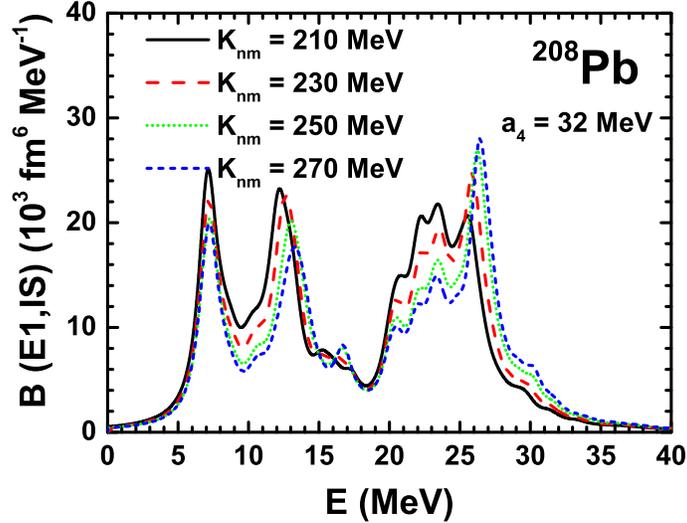}
}
\caption{(Color online) The relativistic RPA isoscalar dipole transition strength in $^{208}$Pb.
The four curves correspond to effective interactions that differ only in the value
of the nuclear matter incompressibility.}
\label{IS-208Pb}
\end{figure}

The isoscalar dipole strength functions for $^{208}$Pb, shown in
Fig.~\ref{IS-208Pb}, have been calculated for a complementary set of
interactions: with the isovector parameters fixed at $a_4 = 32$ MeV
and $L =46.5$ MeV, the four interactions differ only in the value of
the nuclear matter incompressibility: $K_{\rm nm} =$ 210, 230, 250,
and 270 MeV. The high-energy component is obviously very sensitive
to the choice of the nuclear matter incompressibility. Fragmentation
in the region 20 - 30 MeV is somewhat reduced with increasing
$K_{\rm nm}$, and the centroid of this structure is shifted to
higher energies. Varying $K_{\rm nm}$ has very little effect on the
pygmy dipole state: the energy does not change, and only the $B(E1)$
is slightly reduced with increasing incompressibility as the total
strength gets shifted to higher energy. From this result one deduces
that the admixture of $3 \hbar \omega$ compressional mode in the
isoscalar PDS is indeed very small. An intermediate situation is
found for the pronounced resonant structure in the energy region
between 10 and 15 MeV. As a result of the stronger mixing with the
compressional mode the centroid is shifted to higher energy, but at
the same time the transition strength is reduced considerably with
increasing $K_{\rm nm}$.
%
%

\begin{figure}
\centerline{
\includegraphics[scale=0.375,angle=0]{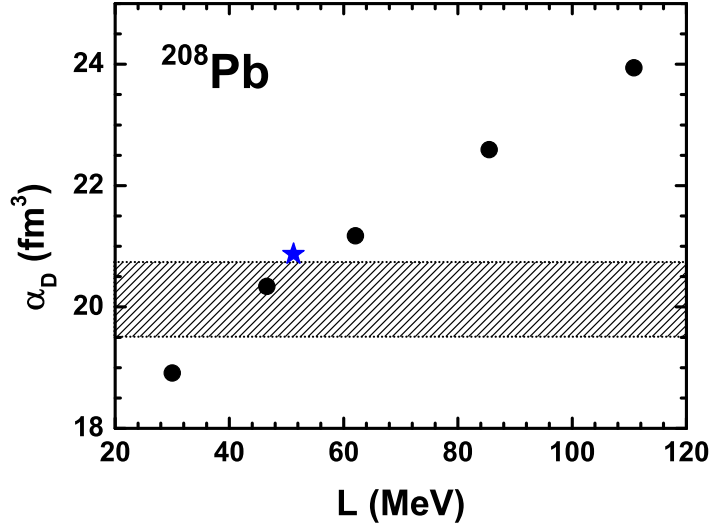}
} \caption{Predictions for the electric dipole polarizability
Eq.~(\ref{dip-pol}) of $^{208}$Pb, calculated with a set of five
effective interactions that differ in the value of the symmetry
energy at saturation ($a_4$) and the corresponding slope parameter
$L$ (circles), and the effective interaction DD-ME2 (star) \cite {LNVR.05}.
The shaded area delineates the experimental constraint $\alpha_D =
(20.1\pm 0.6)$ fm$^3$ \cite{Tam.11}. } \label{dip-pol_208Pb}
\end{figure}

Even though the set of effective interactions that we use here spans a relatively broad
interval of values of the parameters $a_4$ and $L$ that characterize the density
dependence of the symmetry energy, the effect on the excitation energy of the
IV GDR is not large, especially in a heavy nucleus like $^{208}$Pb. In fact, as
shown in Fig.~\ref{E1-IV&IS}, the IV GDR peak energy in $^{208}$Pb is lowered
by less than 1 MeV in the interval between $a_4 = 30$ MeV, $L = 30$ MeV and
$a_4 = 38$ MeV, $L = 111$ MeV. A better distinction between these
effective interactions is obtained by considering predicted values for the
electric dipole polarizability of $^{208}$Pb (cf. Eq.~(\ref{dip-pol})), in
comparison with the experimental value $\alpha_D = (20.1\pm 0.6)$
fm$^3$ \cite{Tam.11}. The results are shown in Fig.~\ref{dip-pol_208Pb}.
In addition to the values obtained with the five effective interactions
$a_4 =$ 30, 32, 34, 36, and 38 MeV, we also include the polarizability calculated
with the interaction DD-ME2 ($a_4 =$ 32.3 MeV and $L = 51.5$ MeV) that
was adjusted independently in Ref.~\cite {LNVR.05}. The result closest to
the experimental value corresponds to the effective interaction with
$a_4 = 32$ MeV and $L = 46.5$ MeV, and the value predicted by DD-ME2:
$\alpha_D = 20.8$ MeV is just 100 keV outside the experimental error bar.
The slope parameters of these two interactions are slightly below the value
$L = 64.8 \pm 15.7$ MeV that Carbone et al. \cite{Car.10} deduced from
the percentage of the EWSR associated with the PDS in $^{68}$Ni \cite{Wie.09}
and $^{132}$Sn \cite{Kli.07}. From Fig.~\ref{dip-pol_208Pb} it is apparent that
only the values of $\alpha_D$ predicted by the two interactions with
$a_4 = 36$ MeV, $L = 85.5$ MeV and $a_4 = 38$ MeV, $L = 110.8$ MeV
are in serious disagreement with experiment.
%
%
\begin{figure}
\centering
\begin{tabular}{c}
\includegraphics[scale=0.3]{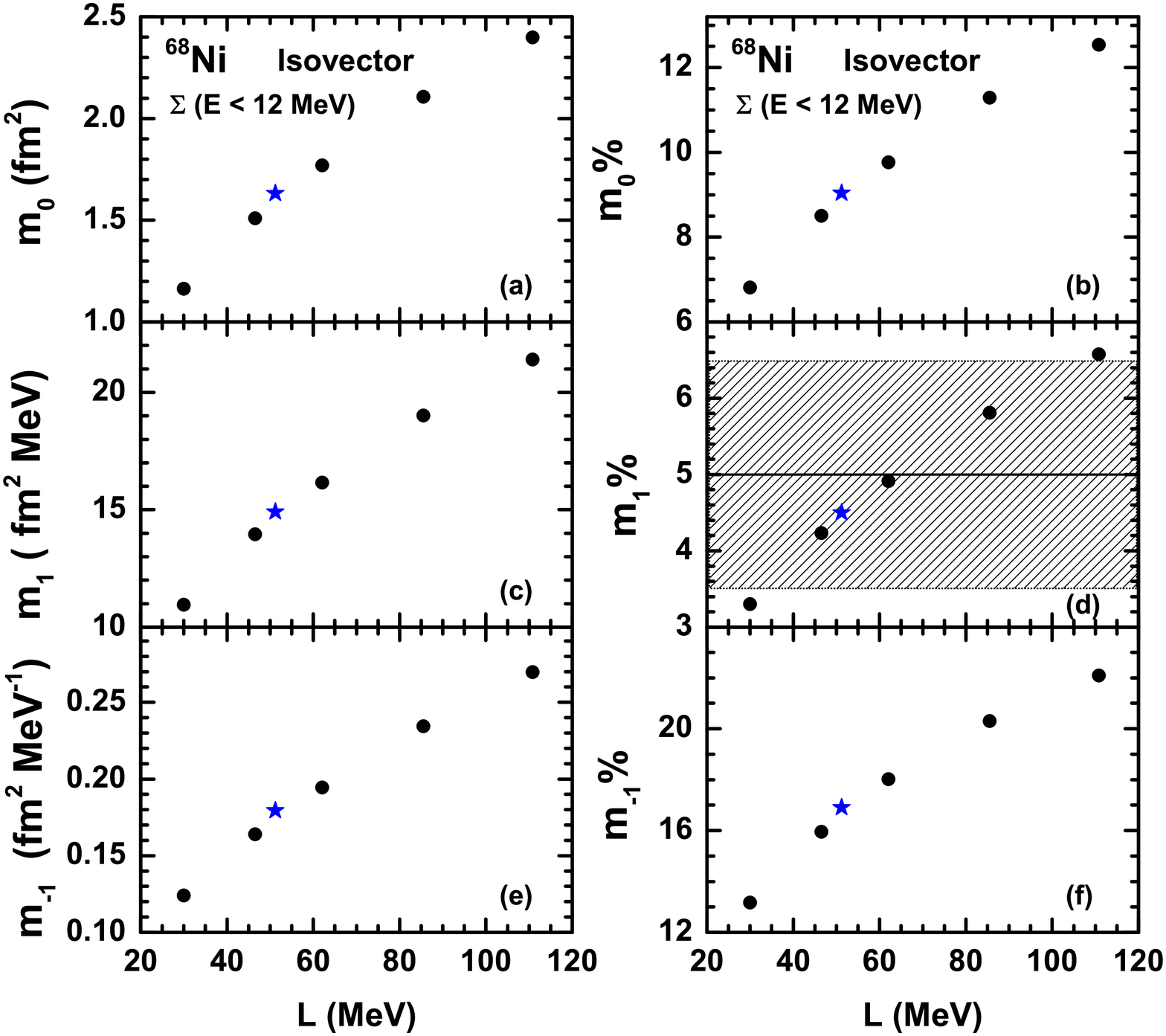} \\
\includegraphics[scale=0.3]{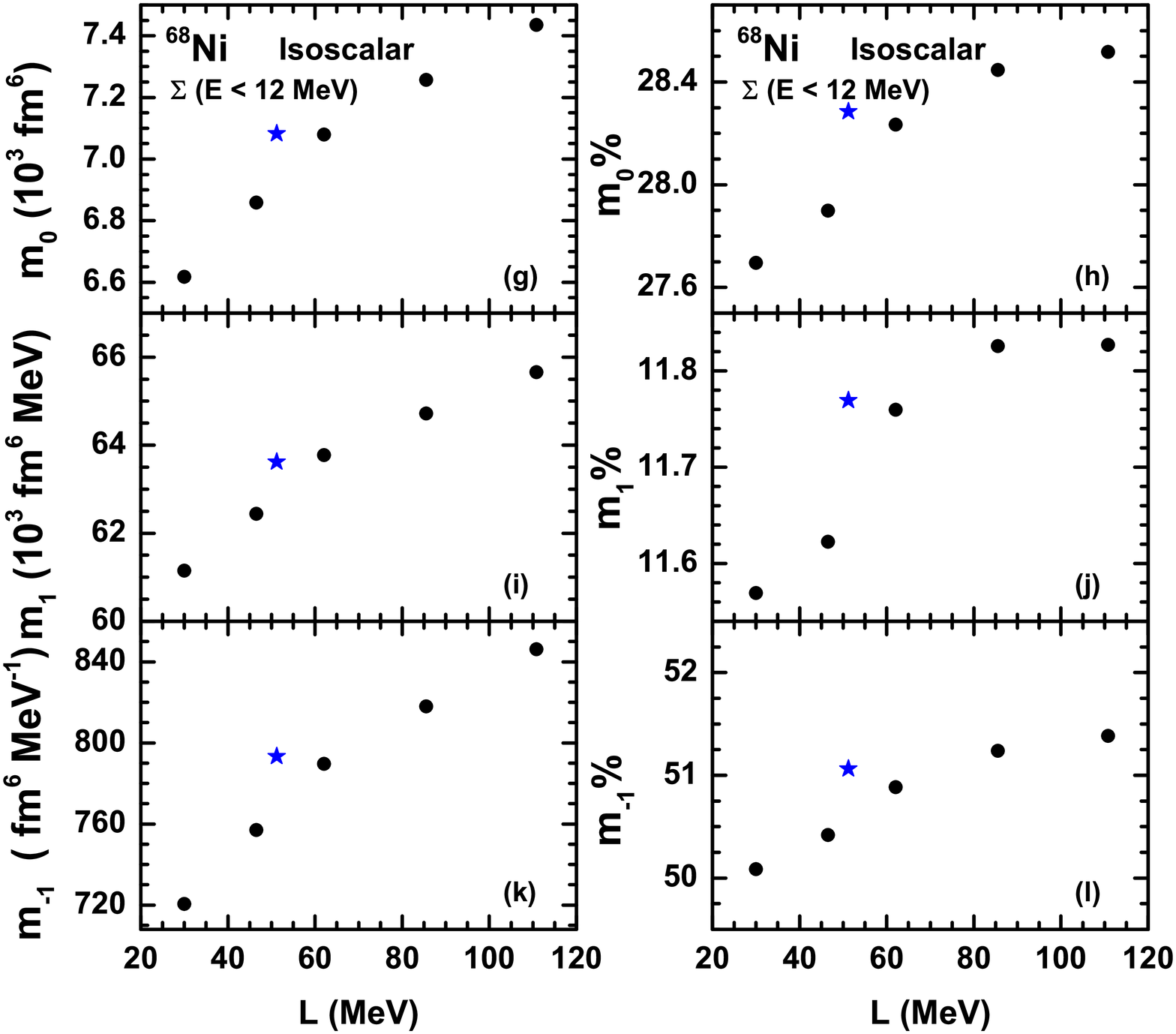}
\end{tabular}
\caption{Moments $m_0$, $m_1$ and $m_{-1}$ of the isovector (panel (a)-(f))
and isoscalar (panel (g)-(i)) dipole strength distributions of $^{68}$Ni. Absolute
values of the moments in the low-energy PDS region below $E < 12$ MeV
(left column), and the percentages of the total moments exhausted by the PDS
(right column), as a function of the slope parameter $L$ of the six effective
interactions used in this work .
}
\label{PDS_moments_Ni}
\end{figure}
%

For $^{68}$Ni in Fig.~\ref{PDS_moments_Ni} we display the moments of the dipole strength
distributions $m_0$, $m_1$ and $m_{-1}$ for the isovector (panel (a)-(f)) and isoscalar
(panel (g)-(i)) channels. $E = 12$ MeV is the model-dependent, although obvious choice
for the excitation energy that separates the low-energy
PDS region from the energy interval dominated by the collective giant resonance
(cf. Fig.~\ref{E1-IV&IS}). In the left column we plot the absolute values of
the moments of the isovector and isoscalar strength distributions in the PDS region, whereas the
column on the right displays the percentages of the total moment exhausted in the PDS interval,
as a function of the slope parameter $L$ of the six effective interactions used in the present analysis.
The shaded area denotes the experimental result and the corresponding uncertainty of
the percentage of EWSR associated with the PDS in $^{68}$Ni \cite{Wie.09,Car.10}.
We note that the absolute values of moments in the low-energy region, as well as
the percentages of the total moment exhausted in the PDS interval, increase linearly with
the slope of the symmetry energy.  In both the isovector and isoscalar channels
the smallest percentage exhausted by the PDS is for the $m_1$ moment. This is because $m_1$ is
energy-weighted and, therefore, dominated by the main giant-resonance structure in
the high-energy region. Much more sensitive to the PDS is the moment $m_{-1}$, directly
proportional to the total polarizability Eq.~(\ref{dip-pol}), because the inverse energy-weighting
enhances the low-energy part of the response. In the isovector channel, in particular, the
percentage of the total polarizability exhausted by the PDS is more than three times
the percentage of the EWSR exhausted by the low-energy pygmy strength. These results
are in agreement with those reported by Piekarewicz \cite{Pie.11}, obtained using
a different class of relativistic effective interactions but with a similar range of values
for the slope parameter of the symmetry energy. What is new here are the results for
the moments of the isoscalar strength distributions shown in the lower panel of
Fig.~\ref{PDS_moments_Ni}. In this case the PDS exhausts more than 10\% of the
EWSR, and more than 50\% of the $m_{-1}$ moment, for all values of the slope
parameter $L$, in agreement with the results of X. Roca-Maza et al.~\cite{RM.11}
obtained using Skyrme effective interactions. Similar results are also found for the
moments of the isovector and isoscalar dipole strength distributions of
$^{132}$Sn and $^{208}$Pb.

An important issue is the degree of collectivity of the PDS. In most
of the early theoretical studies the measure of collectivity was
associated with the number of particle-hole configurations that
significantly contribute to the RPA amplitude of the principal pygmy
state. Such a measure, however, does not take into account the
coherence of these contributions. Lanza et al. \cite{Lan.09}
analyzed the collectivity of PDS by considering the partial
contributions to the reduced transition amplitude. Namely, the
reduced transition probability Eq.~(\ref{BE}) from the ground state
to the excited state $\vert J \lambda \rangle$ can be written as:
\begin{equation}
B^T(EJ) = \left \vert \sum_{ph} A^{T \lambda}_{ph} (EJ) \right \vert^2\; ,
\end{equation}
where the summation is over all particle-hole configurations that
build the excited RPA state. This expression shows that the reduced
transition probability is determined by the number of configurations
that contribute with a significant weight, as well as by the
coherency (relative sign) of these contributions. The RPA state can
be considered as a resonance if the corresponding reduced transition
amplitude is composed of more than just a few particle-hole partial
amplitudes similar in magnitude and with the same relative sign. By
comparing the contributions of the partial amplitudes $A^{T=1
\lambda}_{ph} (E1)$ for the pygmy state and the isovector giant
dipole resonance in $^{132}$Sn, Lanza et al. \cite{Lan.09} concluded
that, although the main pygmy state can indeed be formed by many
particle-hole configurations, the reduced transition probability
$B(E1)$ is generally small because the corresponding partial
amplitudes cancel out to a large extent. This is in contrast to the
GDR for which many configurations add coherently to build a large
transition amplitude. In their study of low-lying dipole response in
$^{68}$Ni, $^{132}$Sn, and $^{208}$Pb, X. Roca-Maza et al.
\cite{RM.11} have found that the isovector reduced amplitude of the
PDS is characterized by destructive interference of a relatively
small number of different particle-hole configurations, and the
resulting reduced transition probability does not exceed $\approx
2-4$ single-particle units. In the isoscalar channel the largest
contributions to the reduced transition amplitude predominantly
arise from neutron particle-hole configurations, and most of them
add coherently. This results in a collective response of the PDS to
the isoscalar dipole operator, characterized by a reduced transition
probability of $\approx 10-20$ single-particle units, for all three
nuclei and the three interactions that were used in the analysis of
Ref.~\cite{RM.11}.

\begin{figure}
\includegraphics[scale=0.325]{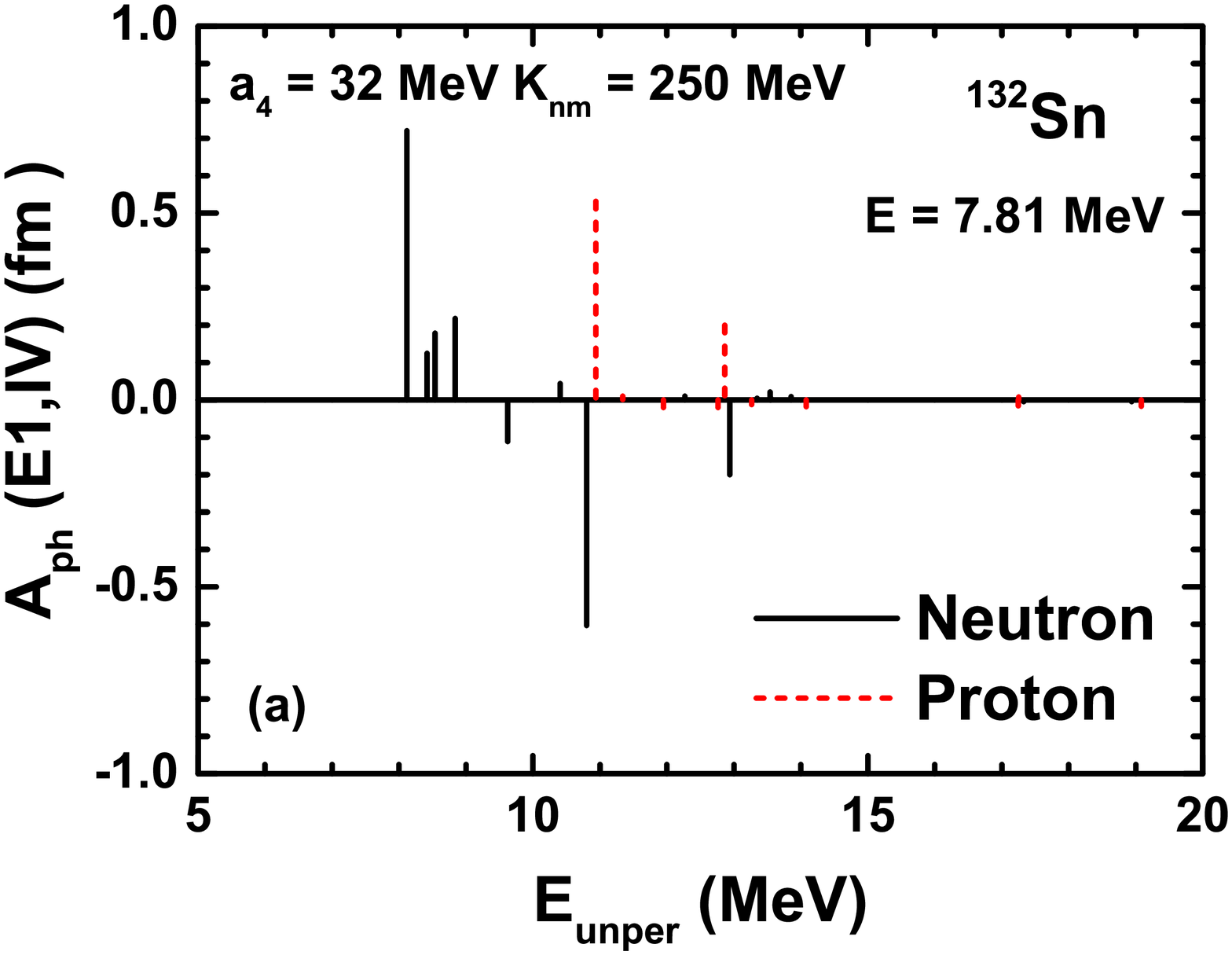}
\includegraphics[scale=0.325]{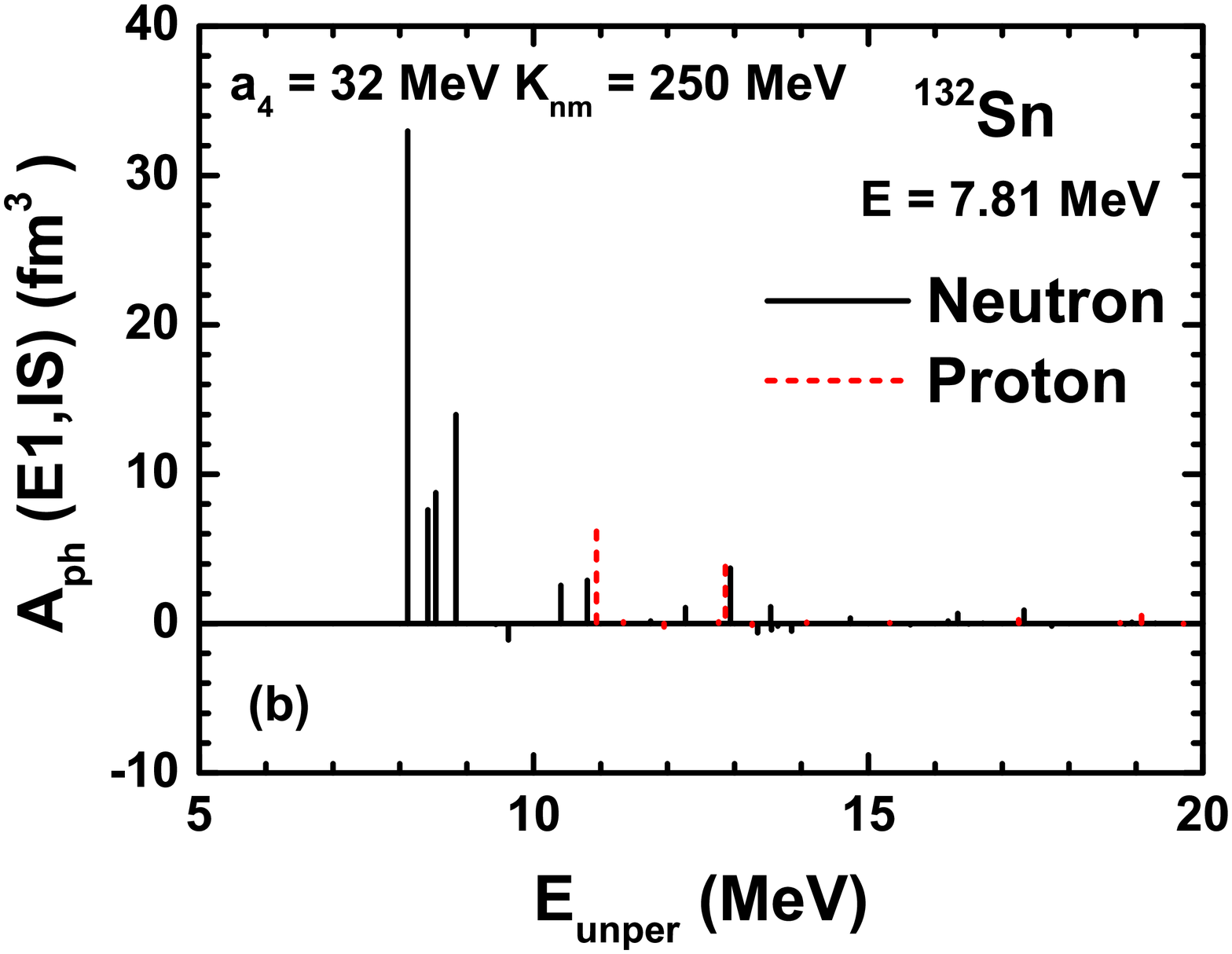}
\caption{(Color online) Partial contributions of neutron and proton
configurations to the isovector (left) and isoscalar (right) reduced
transition amplitude of the PDS state at 7.81 MeV excitation energy
in $^{132}$Sn for the effective
interaction with $a_4=32$ MeV and $K_{\rm nm}=250$ MeV, as
functions of the unperturbed energy of the particle-hole
configurations. Only amplitudes with
a magnitude larger than $\approx 0.01$ fm (isovector), and $\approx 0.1$  fm$^3$
(isoscalar) are shown in the figure.} \label{132Sn}
\end{figure}

\begin{figure}
\includegraphics[scale=0.4]{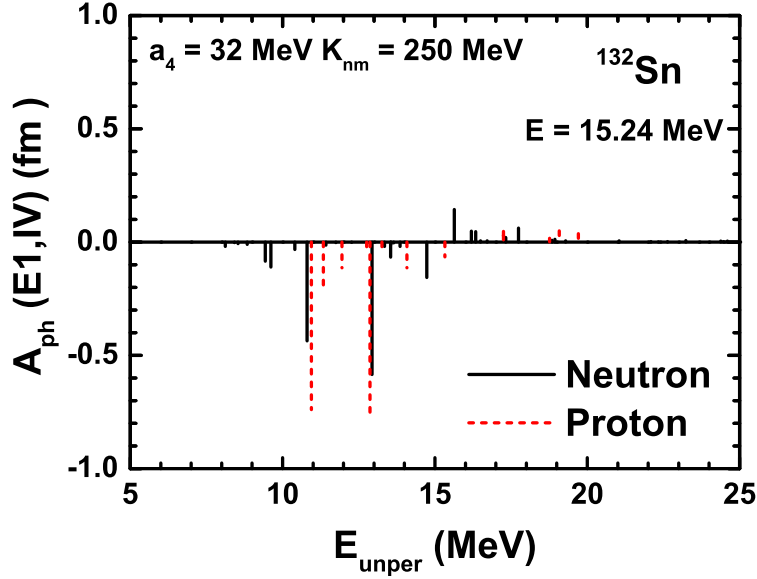}
\caption{(Color online) Partial contributions of neutron and proton
configurations to the isovector GDR state at 15.24 MeV excitation energy
in $^{132}$Sn for the effective
interaction with $a_4=32$ MeV and $K_{\rm nm}=250$ MeV, as
functions of the unperturbed energy of the particle-hole
configurations. Only amplitudes with
a magnitude larger than $\approx 0.01$ fm are shown in the figure.} \label{132Sn_GDR}
\end{figure}

The difference between the PDS in the isovector and isoscalar dipole
response of $^{132}$Sn is illustrated in Fig.~\ref{132Sn}, where we
plot the largest partial contributions of neutron and proton
particle-hole configurations to the isovector (left) and isoscalar
(right) reduced transition amplitude of the main pygmy state at 7.81
MeV excitation energy, calculated using the effective interaction
with $a_4 = 32$ MeV and $L = 46.5$ MeV. The partial reduced
transition amplitudes in units of fm (isovector) and fm$^3$
(isoscalar), are plotted as a function of the unperturbed energy of
the corresponding particle-hole configurations.
Only amplitudes with a magnitude
larger than $\approx 0.01$ fm (isovector), and $\approx 0.1$  fm$^3$
(isoscalar) are shown in the figure. About 7 to 8 neutron
particle-hole configurations display a non-negligible partial
transition amplitude, whereas only two proton configurations
contribute significantly to the isovector and isoscalar amplitudes.
One notices, however, the destructive interference of proton and neutron
particle-hole configurations in the isovector channel (cf. also
Table \ref{Tab_IV}). The dominant neutron amplitudes in the PDS energy region
correspond to the configurations
$3s_{1/2}\rightarrow 3p_{3/2}$, $2d_{3/2}\rightarrow 3p_{1/2}$,
$3s_{1/2}\rightarrow 3p_{1/2}$, and
$2d_{3/2}\rightarrow 3p_{3/2}$, all with positive sign. The two
large negative neutron amplitudes in the GDR region that
correspond to the configurations $1h_{11/2}\rightarrow
1i_{13/2}$ and $1g_{7/2}\rightarrow 1h_{9/2}$, are almost
exactly canceled by the positive proton contributions from
$1g_{9/2}\rightarrow 1h_{11/2}$ and $1f_{5/2}\rightarrow 1g_{7/2}$,
respectively. It is interesting to note that, as shown in Fig.~\ref{132Sn_GDR},
it is precisely the partial transition amplitudes of the
latter two neutron and two proton configurations that
add coherently (here with a negative sign) to produce the large
collective B(E1) of the IVGDR state at 15.24 MeV.  The repulsive residual
interaction, of course, gathers these contributions and shifts them to
higher energy to build the collective GDR. Therefore, while the large
proton and neutron reduced amplitudes in the high-energy region add
coherently to build the GDR, for the PDS state at 7.81 MeV
the same particle-hole configurations with positive
proton and negative neutron contributions cancel each other, so that
the rather small reduced transition amplitude of the PDS state is dominated
by just a few neutron particle-hole configurations (note that the
unperturbed PDS strength is shifted to lower energy by the residual
interaction).

The picture is very different in the isoscalar channel (cf. also Table \ref{Tab_IS}).
 Also in this case the dominant neutron configurations for the PDS are
 $3s_{1/2}\rightarrow 3p_{3/2}$, $2d_{3/2}\rightarrow 3p_{1/2}$
 $3s_{1/2}\rightarrow 3p_{1/2}$, and
$2d_{3/2}\rightarrow 3p_{3/2}$ (positive sign for the partial
 transition amplitude), but here the contributions from $1h_{11/2}\rightarrow 1i_{13/2}$ and
$1g_{7/2}\rightarrow 1h_{9/2}$  are also positive even though an
order of magnitude smaller. Because most neutron configurations add
coherently to the reduced transition amplitude, the total neutron
amplitude $\approx 84$  fm$^3$ is much larger than the corresponding
proton contribution $\approx 17$  fm$^3$. In the response to the isoscalar dipole operator
all major neutron and proton partial transition amplitudes for the PDS state at 7.81 MeV
are of the same sign, and this coherence leads to the large $B(E1)$ value shown in Fig.~\ref{E1-IV&IS}.
%
%
%
%

  \begin{table}[h]
  \centering
  \caption{Partial neutron and proton isovector reduced transition amplitudes
  $A^{T=1}_{ph} (E1)$ for the main pygmy state in $^{132}$Sn. Included in the table are
  only amplitudes with a magnitude larger than about
$0.1$ fm, calculated for the five effective interactions with $a_4
=$ 30, 32, 34, 36, and 38 MeV, and the correspondingly increasing
values of the slope parameter $L$. The total neutron (n) and proton
(p) amplitudes are obtained by summing over all configurations,
including also those not listed in the table.}
  \label{Tab_IV}
    \begin{tabular}{ccccccc}
      \hline\hline
       $a_4$ (MeV) & & 30 & 32 & 34 & 36 & 38 \\
      $E$ (MeV)   & & 7.94  & 7.81 & 7.71 & 7.67  &  7.62 \\

      \hline
                  & config.  & \multicolumn{5}{c}{$A^{T=1}_{ph} $ (fm)} \\
       neutron    & $3s_{1/2}\rightarrow 3p_{3/2}$  & 0.72  &  0.72  &   0.70  &   0.70 &   0.69 \\
                  & $1h_{11/2}\rightarrow 1i_{13/2}$  & -0.59  &  -0.60  &  -0.60  &  -0.62 &  -0.65 \\
                  & $2d_{3/2}\rightarrow 3p_{1/2}$  &  0.20 &   0.22  &   0.23  &   0.24 &   0.25 \\
                  & $1g_{7/2}\rightarrow 1h_{9/2}$  & -0.19 &  -0.20  &  -0.19  &  -0.20 &  -0.20 \\
                  & $3s_{1/2}\rightarrow 3p_{1/2}$  & 0.18  &  0.17  &   0.16 &   0.16 &   0.16 \\
                  & $2d_{3/2}\rightarrow 3p_{3/2}$  & 0.11  &  0.12  &   0.13  &   0.14  &  0.14 \\
                  & $2d_{5/2}\rightarrow 2f_{7/2}$  & -0.12  & -0.11  &  -0.10  &  -0.08 &  -0.05 \\

       n-total& &     0.32    &  0.38   &    0.42  &     0.49   &   0.55   \\
       \hline
       proton     & $1g_{9/2}\rightarrow 1h_{11/2}$  & 0.49  &   0.53  &   0.55 &    0.60  &  0.66 \\
                  & $1f_{5/2}\rightarrow 1g_{7/2}$  & 0.18  &   0.20  &   0.20  &   0.23  &  0.25 \\

       p-total & &   0.53   &    0.64  &    0.70  &     0.81  &    0.91 \\

      \hline\hline
    \end{tabular}
  \end{table}

We have verified that similar results are also obtained with the
other effective interactions used in this work. Tables \ref{Tab_IV}
and \ref{Tab_IS} list the dominant partial reduced transition
amplitudes for the PDS in the isovector and isoscalar channel in
$^{132}$Sn, respectively, calculated for the five effective
interactions with $a_4 =$ 30, 32, 34, 36, and 38 MeV and the
corresponding increasing values of the slope parameter $L$.
In the isovector channel both the neutron and
proton reduced transition amplitudes increase with $a_4$ and $L$ (cf.
Fig.~\ref{E1-IV&IS}). Table \ref{Tab_IV} shows that
the contributions of the neutron and proton particle-hole
configurations increase by 72\% and 70\%, respectively, from $a_4 =
30$ MeV, $L = 30$ MeV to $a_4 = 38$ MeV, $L = 110.8$ MeV. The
increase of $a_4$ and $L$ simply corresponds to an enhancement of
the restoring force for isovector oscillations. However, because of the
partial cancellation of proton and neutron transition amplitudes, the isovector
$B(E1)$ of the PDS is generally small for all values of $a_4$ and $L$. The
occurrence and evolution of strength of the PDS has been
associated with the neutron excess.
The correlation between neutron-skin thickness and the parameters
$a_4$ and $L$ has been studied using a variety of non-relativistic
and relativistic mean-field models \cite{Fur.02,Chen.05,Kli.07}. It
has been shown that for a given nucleus the thickness of the neutron
skin increases linearly with $a_4$ and $L$. In fact, using the same
set of effective interactions, in Ref.~\cite{Kli.07} we have
calculated the increase of the neutron skin in $^{132}$Sn from $r_n
- r_p = 0.19$ fm for $a_4 = 30$ MeV, $L = 30$ MeV to $r_n - r_p =
0.36$ fm for  $a_4 = 38$ MeV, $L = 110.8$ MeV. A number of
studies, including the present, have shown that the strength of the
pygmy state in the isovector response is enhanced with increasing
$a_4$ and $L$, and this enhancement has been correlated with the
corresponding increase of the neutron-skin thickness.
  \begin{table}[h]
  \centering
      \caption{Same as described in the caption to Table \ref{Tab_IV}, but for the
      partial isoscalar reduced transition amplitudes $A^{T=0}_{ph}
      (E1)$ with a magnitude larger than about $1$ fm$^3$.
 }
       \label{Tab_IS}
    \begin{tabular}{ccccccc}
      \hline\hline
         $a_4$ (MeV) & & 30 & 32 & 34 & 36 & 38 \\
       $E$ (MeV)   & & 7.94  & 7.81 & 7.71 & 7.67  &  7.62 \\

      \hline
                  & config.  & \multicolumn{5}{c}{$A^{T=0}_{ph} $ (fm$^3$)} \\
       neutron    & $3s_{1/2}\rightarrow 3p_{3/2}$  & 33.43  &  32.96 &   32.45 &   31.75  &  31.18  \\
                  & $2d_{3/2}\rightarrow 3p_{1/2}$  & 12.85  &  14.13  &  15.08  &  15.98  &  16.48  \\
                  & $3s_{1/2}\rightarrow 3p_{1/2}$  &  9.10  &   8.72  &   8.11  &   7.87  &  7.75\\
                  & $2d_{3/2}\rightarrow 3p_{3/2}$  & 7.11   &  7.73  &   8.41  &   8.75 &   8.78  \\
                  & $1g_{7/2}\rightarrow 1h_{9/2}$  & 3.86   &  3.72  &   3.51  &   3.46 &   3.43 \\
                  & $1h_{11/2}\rightarrow 1i_{13/2}$&  3.24  &   2.88   &  2.45  &   2.27  &   2.18\\
                  & $2d_{5/2}\rightarrow 3p_{3/2}$  & 1.92  &   2.60  &   3.07  &   3.78  &  4.34 \\
                  & $1h_{11/2}\rightarrow 3i_{13/2}$& 1.18  &   1.32  &   1.41  &   1.57  &  1.69 \\
                  & $2d_{5/2}\rightarrow 2f_{7/2}$  & -1.16  &  -1.10  &  -1.11 & -0.86  & -0.57 \\

                  & $1g_{7/2}\rightarrow 2f_{5/2}$  & 0.84 &  1.12  &   1.33  &   1.77  &  2.23 \\
                  & $1h_{11/2}\rightarrow 2g_{9/2}$  & 0.61 & 1.08  &   1.45  &   2.06 &   2.60 \\
                  & $1h_{11/2}\rightarrow 2i_{13/2}$  & 0.81 & 0.91 & 0.98 & 1.08  &  1.15 \\
       n-total & & 79.70 &    83.89  &   86.29   &   90.76  &   94.31 \\

                  \hline
       proton     & $1g_{9/2}\rightarrow 1h_{11/2}$  & 5.36  &  6.19  &   6.68  &    7.73  &   8.74  \\
                  & $1f_{5/2}\rightarrow 1g_{7/2}$  & 3.32  &  3.82  &   4.11 &     4.76  &   5.38  \\
      p-total &  & 15.69  &    16.95 &   17.83   &     19.89 &    21.77 \\
      \hline\hline
    \end{tabular}

\end{table}

The situation is different in the isoscalar response. In this case
the increase of the total proton transition amplitude between $a_4 =
30$ MeV, $L = 30$ MeV and $a_4 = 38$ MeV, $L = 110.8$ MeV (39\%) is
much larger than that of the neutron transition amplitude (18\%).
However, the total contribution of proton configurations to the
overall amplitude is small, varying
from 16\% for $a_4 = 30$ MeV, $L = 30$ MeV, to 19\% for $a_4 = 38$
MeV, $L = 110.8$ MeV. The structure of the PDS in the isoscalar
response is, therefore, dominated by a relatively large number of
neutron particle-hole configurations that, together with
proton configurations, add coherently to build a
reduced transition probability of collective strength. We also
notice that with the systematic increase of $a_4$ and $L$  the
excitation energy of the PDS actually decreases, in contrast to the
result reported in Ref.~\cite{RM.11} but consistent with the
behavior of the corresponding IV GDR (cf. Fig.~\ref{E1-IV&IS}).

\begin{figure}
\includegraphics[scale=0.5]{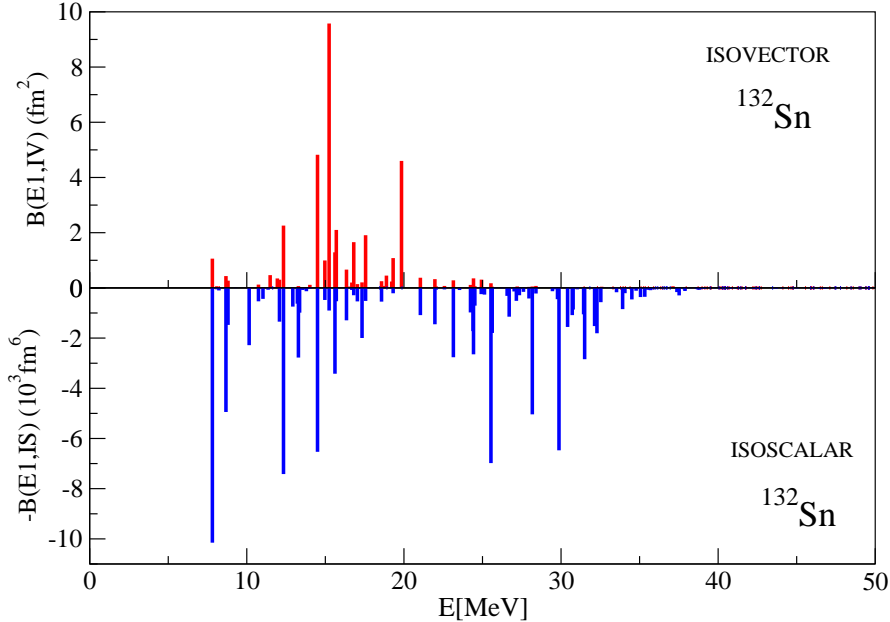}
\caption{(Color online) Discrete $B(E1)$ spectra of the isovector
and isoscalar dipole operators in $^{132}$Sn, calculated using the
effective interaction  with $a_4 = 32$ MeV ($L=46.5$ MeV), and
$K_{\rm nm}=250$ MeV.} \label{Sn132_peaks}
\end{figure}

The discrete isovector and isoscalar $B(E1)$ spectra for $^{132}$Sn
are compared in Fig.~\ref{Sn132_peaks}, for the effective
interaction with $a_4 = 32$ MeV, $L=46.5$ MeV, and $K_{\rm nm}=250$
MeV. At energies below $\approx 20$ MeV many RRPA states display
pronounced  $B(E1)$ values both in the response to the isovector and
isocalar operators. This clearly points to a strong mixing between
the isovector and isoscalar channels, not only for the PDS but also
at higher energies. Several states obviously manifest predominantly
isoscalar or isovector character, e.g. the pygmy state at 7.81 MeV
is mainly isoscalar, whereas the main GDR state at 15.24 MeV is
primarily isovector. At excitation energies above $\approx 20$ MeV
the $B(E1)$ spectra basically contain only isoscalar transitions
that form the broad resonance corresponding to the dipole
compressional mode. The present results are in agreement with those
obtained in a previous RQRPA study of PDS based on transition
densities~\cite{Paa.09}. Theoretical studies have also confirmed
experimental evidence for the splitting of low-energy E1 strength
into two groups of of states with different isospin structure, one
that is excited in $(\alpha,\alpha'\gamma)$ and  $(\gamma,\gamma')$
reactions, and another group of states that is excited only in
$(\gamma,\gamma')$ reactions~\cite{Sav.06,End.09,End.10}. However,
as the present analysis indicates, rather than finding only two
groups of states in the low-energy region, one could actually expect
a more complex pattern of mixed isoscalar and isovector states
across the dipole excitation spectra up to $\approx 20$ MeV.
%
\begin{figure}
\includegraphics[scale=0.6]{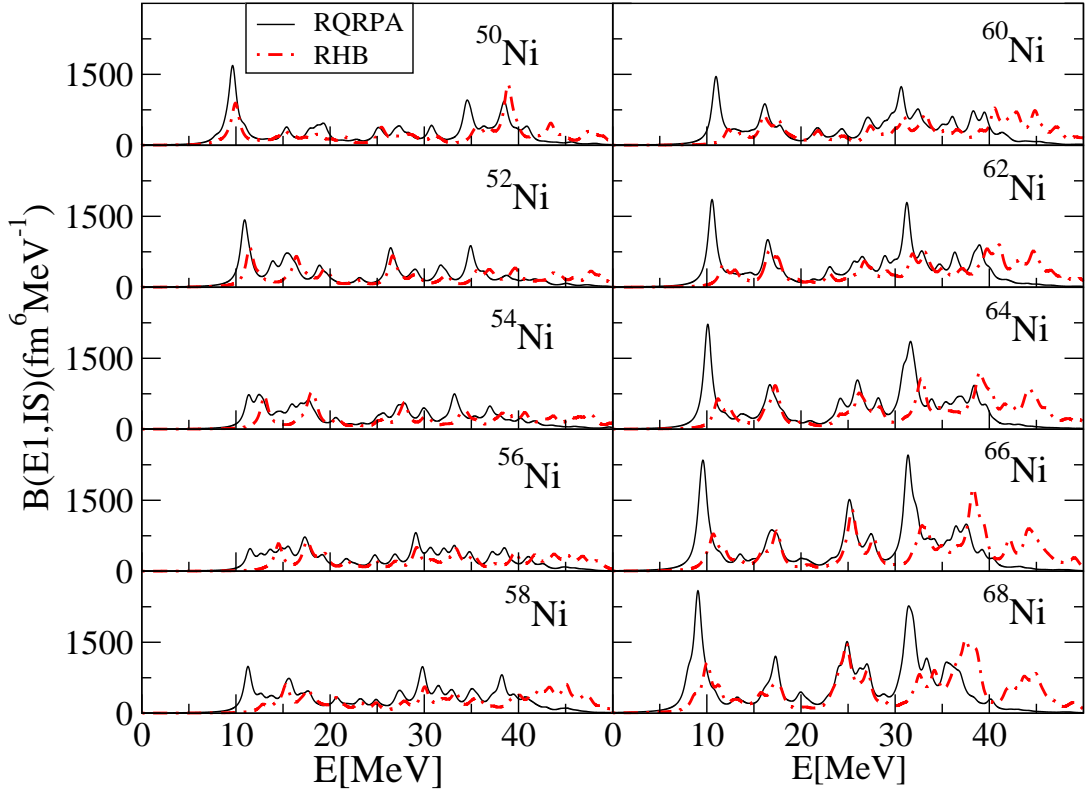}
\caption{(Color online) Evolution of the isoscalar dipole strength distribution in the even-even isotopes
$^{50-68}$Ni, calculated with the RHB+RQRPA model. The DD-ME2 effective interaction is used in the
particle-hole channel, and the Gogny D1S force in the particle-particle channel. The dot-dashed (red)
curve corresponds the unperturbed RHB response.}
\label{Ni_isotopes}
\end{figure}

Finally, to illustrate the role of proton-to-neutron asymmetry in the evolution of PDS in the
isoscalar channel, in Fig.~\ref{Ni_isotopes} we display the isoscalar dipole response for the even-even
isotopes $^{50-68}$Ni. The calculation has been performed with the RQRPA based on the relativistic
Hartree-Bogoliubov (RHB) model, with the DD-ME2 effective interaction in the
particle-hole channel, and the Gogny D1S force in the particle-particle channel. For each nucleus the
RQRPA strength distribution is shown in comparison with the unperturbed RHB response. We find that
for $N \approx Z$ nuclei the transition strength is strongly fragmented, whereas
in other isotopes, either with proton or neutron excess, more pronounced peaks are calculated
both in the low-energy region and at higher energies. The low-energy strength is very
sensitive to the proton-to-neutron asymmetry. The reason is its
underlying nature, i.e. the transition densities for $^{50,52}$Ni are dominated by
proton configurations, and their contribution decreases towards $N \approx Z$.
On the other hand, for $^{62-68}$Ni the PDS is predominantly built from neutron configurations.
Also the high-energy strength around 30 MeV that corresponds to the $3\hbar\omega$ compressional
mode appears to be rather sensitive to the neutron excess. In isotopes beyond $^{60}$Ni
a pronounced peak develops at an excitation energy of $\approx$ §31 MeV. In all isotopes the structure
between the PDS and the high-energy compressional peak essentially coincides with the
unperturbed RHB response.
%
\section{Summary}

In conclusion, we have employed the fully self-consistent random
phase approximation (RPA), based on the framework of relativistic
energy density functionals, to analyze the isovector and isoscalar
dipole response in $^{68}$Ni, $^{132}$Sn, and $^{208}$Pb. In
particular, the evolution of pygmy dipole states (PDS) in the region
of low excitation energies has been analyzed as a function of the
density-dependence of the symmetry energy for a set of relativistic
effective interactions. These interactions, adjusted to empirical
properties of nuclear matter, binding energies and charge radii of
ten spherical nuclei, principally differ in their isovector
properties. They span a broad range of values of the two parameters
that determine the density-dependence of the symmetry energy in
nuclear matter: the symmetry energy at saturation density $a_4$, and
the ``slope" parameter $L$ proportional to the first derivative of
the symmetry energy at saturation. The present study has confirmed
results recently obtained in the framework of non-relativistic and
relativistic mean-field plus RPA models: (i) the range of values of
the slope parameter $L$ constrained by the measured electric dipole
polarizability of $^{208}$Pb \cite{Tam.11} is consistent with the
values deduced from the percentage of EWSR associated with the PDS
in $^{68}$Ni \cite{Wie.09} and $^{132}$Sn \cite{Kli.07}; (ii) the
occurrence of PDS is predicted in the response to both the isovector
and isoscalar dipole operators, and its strength is enhanced with
the increase of $a_4$ and $L$; (iii) in both channels the PDS
exhausts a relatively small fraction of the EWSR but a much larger
percentage of the inverse energy-weighted sum rule ($m_{-1}$ moment
of the strength distribution): $\approx 20\%$ in the isovector
channel and more than 50\% in the isoscalar channel; (iv) for the
isovector dipole operator the reduced transition probability $B(E1)$
of the PDS  is generally small because of pronounced cancellation of
proton and neutron particle-hole contributions to the reduced transition
amplitude; (v) the isoscalar reduced transition amplitude is
predominantly determined by neutron particle-hole configurations,
most of them add coherently and, together with proton contributions, this
results in a collective response of the PDS to the isoscalar dipole operator.
The study of Ref.~\cite{RM.11}, performed with the Skyrme HF+RPA model,
as well as the present analysis based on relativistic effective
interactions, have shown that the PDS is much more pronounced in the
response to the isoscalar dipole operator,  and can be considered as a
resonance state only in the isoscalar channel. This result indicates
that isoscalar probes might be a more appropriate tool for
experimental studies of pygmy dipole states in neutron-rich nuclei
\cite{End.09,End.10,Lan.11,Vit.09}.

\bigskip
\begin{acknowledgements}


This work was supported by the MZOS - project 1191005-1010 and the
Croatian Science Foundation, as well as the Major State 973 Program
2007CB815000, the National Natural Science Foundation of China under
Grants No. 10975008 and 11175002, and the Research Fund for the
Doctoral Program of Higher Education under Grant No. 20110001110087
in China.
\end{acknowledgements}
\clearpage


%

\end{document}